\documentclass[twocolumn,english,superscriptaddress,notitlepage]{revtex4-1}
\usepackage[T1]{fontenc}
\usepackage[latin9]{inputenc}
\usepackage{graphicx}
\usepackage{amsmath}
\usepackage[usenames, dvipsnames]{color}

\makeatletter
\usepackage{babel}
\usepackage{float}

\makeatother

\begin{document}

\title{Homogenous reduced moment in a gapful scalar chiral kagome antiferromagnet}

\author{A. Scheie}
\address{Institute for Quantum Matter and Department of Physics and Astronomy, Johns Hopkins University, Baltimore, MD 21218}

\author{S. Dasgupta }
\address{Institute for Quantum Matter and Department of Physics and Astronomy, Johns Hopkins University, Baltimore, MD 21218}

\author{M. Sanders}
\address{Department of Chemistry, Princeton University, Princeton, NJ 08544 }

\author{A. Sakai}
\address{Institute for Solid State Physics, University of Tokyo, Kashiwa, Chiba 277-8581, Japan}

\author{Y. Matsumoto}
\address{Institute for Solid State Physics, University of Tokyo, Kashiwa, Chiba 277-8581, Japan}

\author{T.R. Prisk}
\address{NIST Center for Neutron Research, National Institute of Standards and Technology, Gaithersburg, MD 20899}

\author{S. Nakatsuji}
\address{Institute for Solid State Physics, University of Tokyo, Kashiwa, Chiba 277-8581, Japan}

\author{R.J. Cava}
\address{Department of Chemistry, Princeton University, Princeton, NJ 08544 }


\author{C. Broholm}
\address{Institute for Quantum Matter and Department of Physics and Astronomy, Johns Hopkins University, Baltimore, MD 21218}
\address{NIST Center for Neutron Research, National Institute of Standards and Technology, Gaithersburg, MD 20899}
\address{Department of Materials Science and Engineering, Johns Hopkins University, Baltimore, MD 21218}

\date{\today}

\begin{abstract}
		We present present a quantitative experimental investigation of the scalar chiral magnetic order with in $\rm{Nd_3Sb_3Mg_2O_{14}}$. Static magnetization reveals a net ferromagnetic ground state, and inelastic neutron scattering from the hyperfine coupled nuclear spin reveals a local ordered moment of 1.76(6) $\mu_B$, just 61(2)\% of the saturated moment size. The experiments exclude static disorder as the source of the reduced moment. A 38(1)$\> \mu$eV gap in the magnetic excitation spectrum inferred from heat capacity rules out thermal fluctuations and suggests a multipolar explanation for the moment reduction. We compare $\rm{Nd_3Sb_3Mg_2O_{14}}$ to Nd pyrochlores and show that it is close to a moment fragmented state.
\end{abstract}

\maketitle
	
\section{Introduction}

A new family of rare earth kagome compounds $\rm{RE_3Sb_3A_2O_{14}}$ (RE = rare earth, A = Mg, Zn) has recently been discovered \cite{SandersREMg,SandersREZn,Dun2016,Dun2017,MyPaper}. These materials, sometimes called "tripod kagome", host a variety of magnetic phases, including topological scalar chiral order \cite{MyPaper}, emergent charge order \cite{Paddison2016}, quantum spin fragmentation \cite{dun2018quantum}, and a quantum spin liquid phase \cite{Ding2018}.
To date, the magnetic structures of three of these compounds ($\rm{Nd_3Sb_3Mg_2O_{14}}$,  $\rm{Dy_3Sb_3Mg_2O_{14}}$, and  $\rm{Ho_3Sb_3Mg_2O_{14}}$) have been determined by powder neutron diffraction \cite{MyPaper,Paddison2016,dun2018quantum} and found to share two characteristics: an average all-in-all-out (AIAO) order (where the ordered spins point into or out of a triangle center) with a net ferromagnetic moment along the $c$ axis, and an ordered magnetic moment significantly below the saturated moment expected for the magnetic ion.
The ferromagnetic AIAO order  is interesting because it indicates a net scalar chirality (where the scalar triple product of three spins around a triangle ${\bf S}_1 \cdot ({\bf S}_2 \times {\bf S}_3) \neq 0 $) and topologically protected magnon edge states \cite{Hirschenberger2015,Owerre2017,Laurell_2018}.
The reduced ordered moment, meanwhile, seems to indicate a disordered or fluctuating ground state. In $\rm{Dy_3Sb_3Mg_2O_{14}}$ and  $\rm{Ho_3Sb_3Mg_2O_{14}}$ it has been proposed, based on elastic diffuse neutron scattering, that the reduced moment results primarily from static spin disorder in an emergent magnetic charge ordered two-in-one-out two-out-one-in state \cite{Paddison2016,dun2018quantum}. 

In this paper we (i) use low temperature static magnetization to explicitly show there is a net ferromagnetic component of the magnetic order in $\rm{Nd_3Sb_3Mg_2O_{14}}$ as previously inferred from neutron diffraction. (ii) Through neutron measurements of nuclear hyperfine splitting, we show there is a uniform 40\% reduction of the ordered moment per site relative to the saturation moment.  (iii) We show there is a 40 $\mu$eV gap in the magnetic excitation spectrum through analysis of the low $T$ specific heat. These results lead to a broader discussion of reasons behind and methods to detect moment reduction in frustrated rare earth based magnets.

\section{Experiments}

\begin{figure}
	\centering\includegraphics[scale=0.45]{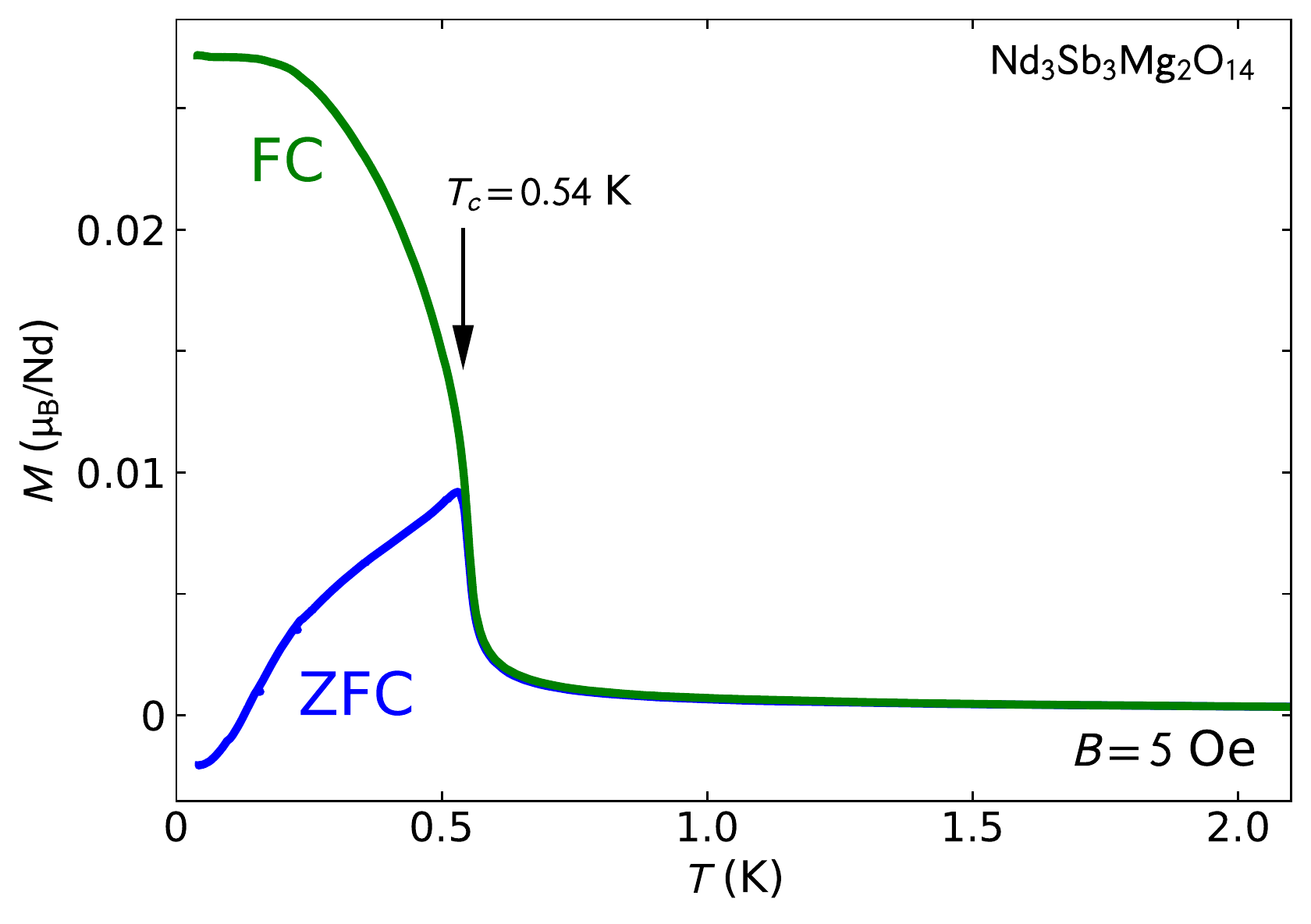}
	
	\caption{Low temperature powder-averaged magnetization of $\rm{Nd_3Sb_3Mg_2O_{14}}$. Measurements were taken upon heating from a field-cooled (green) and zero-field-cooled (blue) state. The splitting of these two curves indicates a ferromagnetic ground state.}
	\label{flo:magnetization}
	
\end{figure}

\paragraph{Magnetization:}
We measured low temperature magnetization of 0.1 mg loose powder $\rm{Nd_3Sb_3Mg_2O_{14}}$ using a custom-built SQUID magnetometer. The loose powder was mixed with silver paste and attached to silver foil to ensure good thermal connection. We measured static magnetization as a function of temperature from 40 mK to 2.2 K at 5 Oe twice: once starting from a field-cooled (FC) and again from a zero-field-cooled (ZFC) state. The data is shown in Fig. \ref{flo:magnetization}.
The low temperature SQUID magnetometer only measures relative magnetization, so we normalized the data to units of $\rm \mu_B / Nd$ by scaling the low temperature SQUID data from 1.8 K to 2.2 K to match magnetization data taken on an MPMS (31.3 mg, no silver powder, 800 Oe) \cite{NIST_disclaimer}. 

\paragraph{Neutron Scattering:}

We measured neutron scattering on 20.3 g loose powder  $\rm{Nd_3Sb_3Mg_2O_{14}}$ using the HFBS backscattering spectrometer at the NCNR. The sample was sealed under 10 bar helium in a copper can that was attached to the mixing chamber of a dilution refrigerator. We measured at a bandwidth of $\pm 36 \> {\rm \mu eV}$ and an elastic full width at half maximum  energy resolution of 1.04 $\rm \mu eV$ at 50 mK for 21.5 hours, 700 mK for 4 hours, and at 4.5 K for 2 hours. To more accurately measure the hyperfine excitations, we switched to the $\pm 11 \> {\rm \mu eV}$ mode (0.79 $\rm \mu eV$ resolution), measuring for six hours at base temperature, two hours at 300 mK, and for two hours at 700 mK. The data measured in the $\pm 11 \> {\rm \mu eV}$ mode are shown in Fig. \ref{flo:hyperfine}. The $\pm 36 \> {\rm \mu eV}$ data (in which no spin waves are visible) is given in the supplemental information.

The high energy resolution of the HFBS spectrometer allows detection of nuclear spin flip excitations in the hyperfine enhanced field associated with the 4f electronic dipole moments.  The corresponding scattering cross section takes the form of a low-energy peak at the nuclear spin flip energy with an intensity that is $Q$-independent (except for the Debye Waller factor). This scattering is distinguished from magnetic inelastic scattering which is typically dispersive with an intensity that follows the electronic magnetic form factor  \cite{HTO_nuclearmagnetism,przenioslo2006nuclear}. Nd has two isotopes with nuclear moments: 12.2\% Nd$^{143}$ (incoherent cross section $\sigma_i = 55(7) \>$barn) and 8.29\% Nd$^{145}$ ($\sigma_i = 5(5) \>$barn), both with nuclear spin $I=9/2$. When $\rm{Nd_3Sb_3Mg_2O_{14}}$ orders below 540 mK, the nuclear spin levels will be split, and we expect to see nuclear hyperfine excitations.

\begin{figure}
	\centering\includegraphics[scale=0.55]{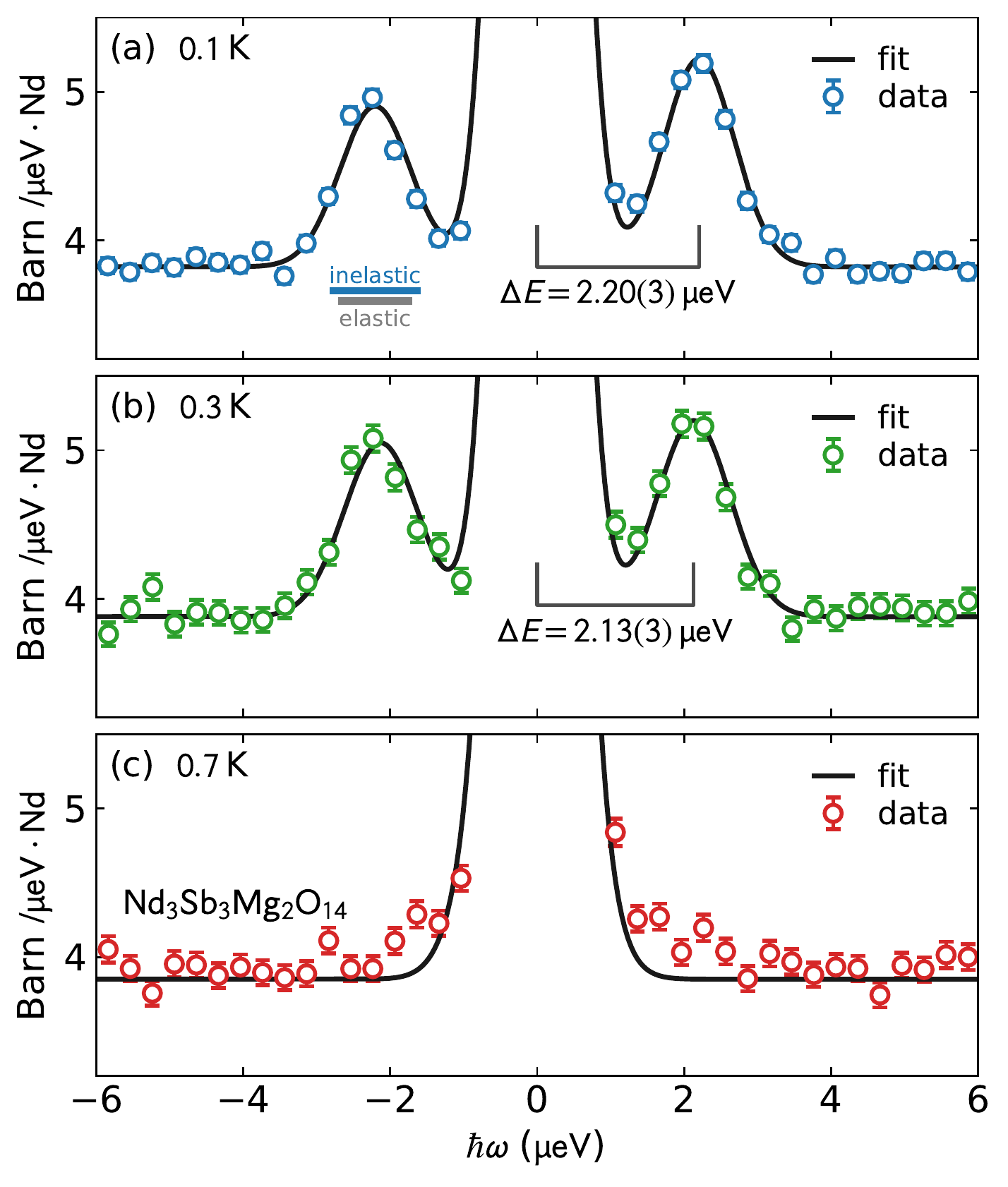}
	
	\caption{Hyperfine nuclear excitations in $\rm{Nd_3Sb_3Mg_2O_{14}}$ measured by neutron scattering at 0.1 K, 0.3 K, and 0.7 K. These data are the sum over all detectors (0.25 \AA$^{-1}$ to 1.75 \AA$^{-1}$). Each nuclear excitation peak is slightly wider than the resolution width as shown in panel (a). The energy of the excitation peaks indicate the size of the ordered electronic magnetic moment. There is an unknown temperature independent offset of the measured intensity from zero arising from background contributions to the detector count rate. Error bars represent one standard deviation.}
	\label{flo:hyperfine}
	
\end{figure}

The neutron cross section of powder-averaged nuclear hyperfine excitations is
\begin{equation}
\Big( \frac{d^2 \sigma}{d \Omega d E} \Big)^{\pm} = \frac{1}{3} \frac{k_f}{k_i} e^{-2W(Q)} \overline{I(I+1) \frac{\sigma_i}{4 \pi}
	\delta(\Delta_M - E)}, 
\end{equation}
where $\pm$ refers to positive and negative energy transfer, $\Delta_M$ is the hyperfine splitting energy, $\sigma_i$ is the incoherent scattering cross section for the magnetic ion, $I$ is the nuclear spin state, $2W(Q)=\langle u^2 \rangle Q^2$ and $\langle u^2 \rangle$ is the mean squared displacement of the nucleus, $k_i$ and $k_f$ are the incident and scattered neutron wave vectors, and the horizontal bar indicates an isotope average  \cite{heidemann1970}. For Nd, $\overline{I(I+1) \frac{\sigma_i}{4 \pi}} = 8.98 \> \>{\rm barn}$. Using eq. 1, we were able to determine the energies of the nuclear hyperfine excitations and use the hyperfine integrated intensity to convert the data to absolute units.

\section{Results}

\paragraph{Magnetization:}

The bifurcation between the FC and ZFC magnetization measurements in Fig. \ref{flo:magnetization} clearly indicate a ferromagnetic transition at $T_c = 0.54 \> {\rm K}$.
When a ferromagnet orders in zero field, the domains form with random orientations resulting in a net zero magnetization. When cooled in field, the ferromagnetic domains preferentially order along the field direction, giving a non-zero magnetization. Thus, a key signature of a ferromagnetic material is a difference between the field-cooled and zero-field-cooled magnetization
---precisely what we observe in $\rm{Nd_3Sb_3Mg_2O_{14}}$. In the ZFC data, the normalized magnetization dips slightly below zero. This negative value can be neglected, as it is within the systematic error bars for in the normalization to MPMS data (which can have have nonlinear effects below 0.5 K). Therefore, we confirm the prediction from previous neutron scattering work \cite{MyPaper} that the canted antiferromagnetic order of $\rm{Nd_3Sb_3Mg_2O_{14}}$ has a net ferromagnetism.$\rm{Dy_3Sb_3Mg_2O_{14}}$ and  $\rm{Ho_3Sb_3Mg_2O_{14}}$ have also been inferred to have a ferromagnetic moment based on analysis of the antiferromagnetic diffraction
\cite{Paddison2016,dun2018quantum}.

\paragraph{Hyperfine Excitations:}

The neutron scattering results in Fig. \ref{flo:hyperfine} show the appearance of finite energy nuclear spin flip excitations below  $T_c$. The nuclear hyperfine coupling is too weak to influence the spin dynamics in this system, but it can be used to calculate the local electronic ordered moment.
To extract precise values for the energies, we fit the data with Gaussian peaks weighted by a population factor $e^{\pm \beta \hbar \omega/2}/(e^{-\beta \hbar \omega/2} + e^{\beta \hbar \omega/2})$ as shown in Fig. \ref{flo:hyperfine}. (The temperature was treated as a fitted parameter for the lowest temperature data in Fig. \ref{flo:hyperfine}(a), giving a value of 0.10(3) K. For higher temperature data, $T$ was determined by resistive thermometry.) The 0.1 K data shows an excitation energy of 2.20(3) $\rm \mu eV$, and the 0.3 K data shows an excitation energy of 2.13(3) $\rm \mu eV$. At 0.7 K no nuclear hyperfine excitations are visible, indicating no static electronic moment.
Using the empirical relation (extracted from multiple neutron diffraction and hyperfine experiments on Nd-based magnetic materials) between Nd nuclear hyperfine energies $\Delta E$ and static magnetic moment $\mu$ in ref \cite{Chatterji2008},
\begin{equation}
\Delta E = \mu \times (1.25 \pm 0.04) {\rm \frac{\mu eV}{\mu_B}},
\end{equation}
we calculate a mean ordered Nd moment of 1.76(6) $\rm \mu_B$ at 0.1 K, and 1.70(6) $\rm \mu_B$ at 0.3 K. The hyperfine peaks are slightly wider than the central elastic peak: FWHM = 1.131(6) $\rm \mu eV$ (inelastic 0.1 K) and FWHM = 1.178(7) $\rm \mu eV$ (inelastic 0.3 K) vs FWHM = 0.9059(1) $\rm \mu eV$ (central elastic). This evidences either a finite relaxation rate or a distribution of ordered moments in the sample: $\pm 0.19(2) \> \rm \mu_B$ at 0.1 K, $\pm 0.23(3) \> \rm \mu_B$ at 0.3 K or  (see the supplemental materials for details). 

It is worth emphasizing that these nuclear hyperfine measurements are \textit{local probes} of the Nd magnetism: the hyperfine excitation energy is proportional only to moment size and is independent of the number of atoms involved. In contrast, magnetization and neutron diffraction are extensive quantities that vary in proportion to the sample mass. Although there is a small distribution of ordered moments (from $1.99 \> \rm \mu_B$ to  $1.53 \> \rm \mu_B$) the order is nearly homogeneous with all spins between 1/2 and 2/3 the expected ordered moment.

\begin{table}
	\caption{Low temperature ordered moment of $\rm{Nd_3Sb_3Mg_2O_{14}}$ measured by neutron diffraction, hyperfine excitations, nuclear Schottky anomaly, and calculated the CEF Hamiltonian. The experimental values agree with each other, but not the theoretical value.}
	\begin{ruledtabular}
		\begin{tabular}{cccc}
			Neutron & Hyperfine & Nuclear & CEF \tabularnewline
			diffraction & excitations & Schottky & Theory \tabularnewline
			\hline 
			1.79(5) $\rm \mu_B$  & 1.76(6) $\rm \mu_B$ & 1.73(4)  $\rm \mu_B$ &  2.89 $\rm \mu_B$ \tabularnewline
	\end{tabular}\end{ruledtabular}
	\label{flo:orderedmoment}
\end{table}

The mean hyperfine ordered magnetic moment agrees to within the experimental uncertainty with the ordered moment measured by neutron diffraction: 1.79(5) $\rm \mu_B$ \cite{MyPaper}. As shown in Table \ref{flo:orderedmoment}, the measurements of the ordered moment are 38\% less than the theoretical ordered moment for $\rm{Nd_3Sb_3Mg_2O_{14}}$ calculated from the crystal electric field (CEF) Hamiltonian: 2.89 $\rm \mu_B$ \cite{Scheie2018_CEF} (which fully takes into account atomic scale anisotropies and quantum effects). 

This remarkable agreement between a local probe (hyperfine excitations) and a bulk probe (neutron scattering) of the ordered moment means that the moment reduction cannot arise from any static disorder, as in $\rm{Dy_3Sb_3Mg_2O_{14}}$. 
And yet the measured ordered moment is only 2/3 the saturation moment for the Kramers doublet. This indicates an ordered state that incorporates rather strong quantum fluctuations as might be expected for a frustrated spin system in two dimensions, potentially involving higher-order magnetic order that is invisible to neutrons and hyperfine coupling like in ref. \cite{Benton_2016} (e.g., order in the octupolar level).

\paragraph{Heat Capacity Fits:}

The point group symmetry for Nd$^{3+}$ in $\rm{Nd_3Sb_3Mg_2O_{14}}$ is $2/m$ corresponding to a ligand environment with a strong easy-axis character \cite{Scheie2018_CEF}. Absent an accidental degeneracy, it would be surprising if gapless spin excitations existed in $\rm{Nd_3Sb_3Mg_2O_{14}}$ which produce the reduced ordered moment.

We can determine the spin excitation gap by fitting the low temperature heat capacity from ref. \cite{MyPaper} assuming a gapped bosonic (spin-wave) spectrum:
\begin{equation}
U = \frac{v_0}{(2 \pi)^3} 4 \pi \int \epsilon(q) \frac{1}{e^{\epsilon(q)/k_bT}-1}q^2 dq .
\label{eq:HeatCapacityGap}
\end{equation}
Here $\epsilon(q) = \sqrt{\Delta^2 + (cq)^2}$ is the spin-wave dispersion  with velocity $c$ and gap $\Delta$, and heat capacity is computed as $C = \frac{\partial U}{\partial T}$. We solved these equations numerically, added a Nd nuclear Schottky anomaly \cite{PyNuclearSchottky} so that $C_{tot} = C_{electronic} + C_{nuclear}$, and fit the resulting $C_{tot}$ to the data. The fits are shown in Fig. \ref{flo:HC} (details are given in the supplemental information). The fitted gap is $\Delta=38 \pm 1.4 \> \mu$eV, which is consistent with the absence of magnetic excitations in the neutron scattering data within a bandwidth of $\pm 36 \> \mu$eV. There is in fact evidence for this gap even in the higher temperature heat capacity data from the $x$ intercept in a plot of $C/T$ vs $T^2$ plot [see Fig. \ref{flo:HC}(c)]. Specifically, 
\begin{equation}
\Delta = \left(\sqrt{\frac{8}{5}}\pi k_{B}\right)\sqrt{x_{i}}.
\end{equation}
where $x_i$ is the $x$ intercept of a high temperature linear extrapolation in K$^2$ and $\Delta$ is the gap in meV. (The derivation is given in the supplemental information.) This relation and holds for bosonic quasiparticles with a dispersion relation that can be approximated by the relativistic form, and allows for determination of a gap from data above the gap temperature scale. 

The fitted spin wave velocity is $c = 46.31 \pm 0.08 \> {\rm m/s}$, and a fitted ordered moment is $1.73 \pm 0.04 \> {\rm \mu_B}$. This ordered moment agrees to within uncertainty with the neutron values in Table \ref{flo:orderedmoment} and again indicates a significant moment reduction. 
The existence of this gap means that the reduced ordered moment cannot be from low-lying spin wave states, because they would be depopulated at the lowest temperatures. A possible explanation for the reduced ordered magnetic moment is spin order in a rotated basis involving higher multipoles, as was calculated for $\rm Nd_2Zr_2O_7$ \cite{Benton_2016}, as neutron scattering and hyperfine splitting are both only sensitive to the ordered dipole moment (see supplemental materials for details, which includes refs. \cite{PyCrystalField,Santini_2009,Hermele_2014,Li_2016}).

\begin{figure}
	\centering\includegraphics[scale=0.5]{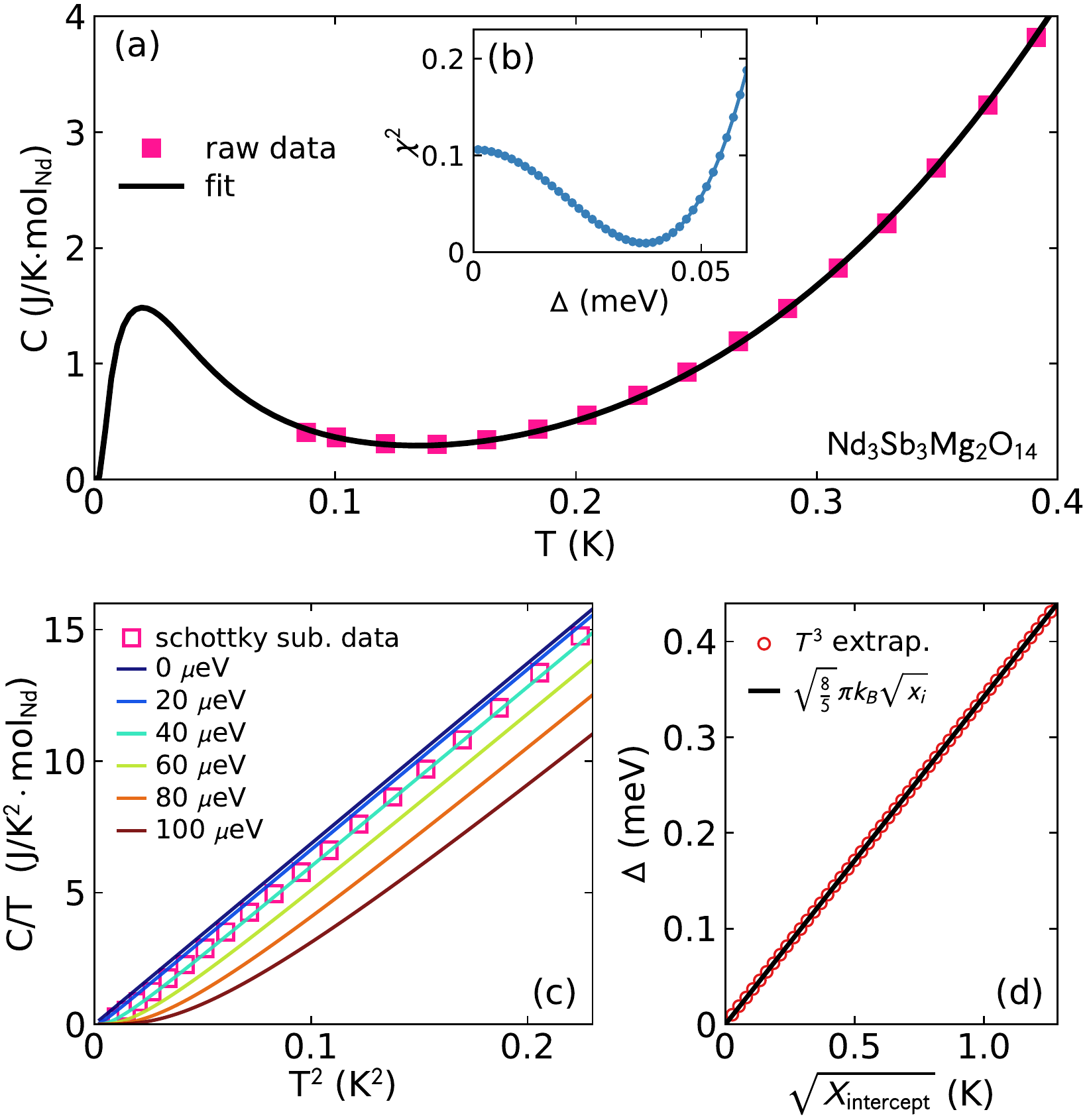}
	
	\caption{Low temperature heat capacity of $\rm{Nd_3Sb_3Mg_2O_{14}}$  fitted to a gapped spin wave spectrum. (a) Heat capacity from ref. \cite{MyPaper} with fit based on eq. \ref{eq:HeatCapacityGap} plus a nuclear Schottky anomaly. (b) Best fit $\chi^2$ as a function of gap energy, showing a minimum at 38(1) $\mu$eV. (c) Data with the fitted nuclear Schottky term subtracted, revealing high temperature $T^3$ behavior with the $x$ intercept determined by the gap size. (d) Gap size plotted against extrapolated $x$ intercept from panel (c), revealing a perfect square root relationship.}
	\label{flo:HC}
\end{figure}

\section{Discussion}

The tripod kagome compounds may be described as 2D versions of pyrochlores compounds. Theory suggests \cite{Dun2016}, and experiments confirm that they often exhibit similar magnetic properties of their pyrochlore parents. For example, $\rm{Dy_3Sb_3Mg_2O_{14}}$ and  $\rm{Ho_3Sb_3Mg_2O_{14}}$ both exhibit kagome-ice magnetic ground states \cite{Paddison2016,dun2018quantum}, like 2D versions of classical spin ices $\rm Dy_2Ti_2O_7$ and $\rm Ho_2Ti_2O_7$ \cite{gingras2011spin}. The same correspondence exists for $\rm{Nd_3Sb_3Mg_2O_{14}}$: the Nd$^{3+}$ pyrochlore compounds $\rm Nd_2Sn_2O_7$, $\rm Nd_2Hf_2O_7$, and $\rm Nd_2Zr_2O_7$ all show AIAO magnetic order with strongly reduced magnetic ordered moments \cite{Anand_2015,Anand_2017,Bertin_2015,Hatnean_2015,Lhotel_2015,Petit2016}.

The similarity is particular striking between the tripod kagome system  $\rm{Nd_3Sb_3Mg_2O_{14}}$ and the pyrochlore $\rm Nd_2Sn_2O_7$, which has a local ordered moment of 1.7 $\mu_B$/Nd (measured by both nuclear hyperfine and neutron diffraction) \cite{Bertin_2015}. Meanwhile, the pyrochlores $\rm Nd_2Hf_2O_7$ and $\rm Nd_2Zr_2O_7$ show more dramatically reduced moments: 0.62(1) $\mu_B$/Nd \cite{Anand_2015} for $\rm Nd_2Hf_2O_7$ and 0.80(5)  $\mu_B$/Nd \cite{Lhotel_2015,Petit2016} or 1.26(2) $\mu_B$/Nd \cite{Xu_2015} (depending on the sample used) for $\rm Nd_2Zr_2O_7$. This massive reduction suggests an additional mechanism behind the moment reduction in $\rm Nd_2Hf_2O_7$ and $\rm Nd_2Zr_2O_7$.

Petit et al. have suggested that $\rm Nd_2Zr_2O_7$ is a "fragmented" spin ice  \cite{Petit2016}: wherein emergent magnetic monopoles order in a long range pattern, forming a three-in-one-out three-out-one-in order on teach tetrahedra. This creates an average AIAO order with a 50\% reduced net magnetic moment \cite{Brooks-Bartlett_2014}. In this way, a spin is "fragmented": part of each spin contributes to a long range pattern but part contributes to a short-range pattern.
The main evidence for this in $\rm Nd_2Zr_2O_7$ is a spin-ice like pinch point neutron spectrum at finite energy. However, ref. \cite{Benton_2016} showed these experimental features can exist without a fragmented spin-ice state involving ordered octupolar moments. 
A moment fragmented state would feature a local ordered moment much larger than the spatial-average moment measured by neutron diffraction, and this remains to be demonstrated.


%

Assuming the moment fragmentation hypothesis is correct, the mere substitution of Sn for Zr changes conventional ordered $\rm Nd_2Sn_2O_7$ to moment fragmented $\rm Nd_2Zr_2O_7$. Given the similarities between $\rm{Nd_3Sb_3Mg_2O_{14}}$ and $\rm Nd_2Sn_2O_7$, this suggests that if appropriate ions could be substituted, $\rm{Nd_3Sb_3Mg_2O_{14}}$ may be driven to a fragmented, fluctuating ground state.

\section{Conclusion}

In conclusion, we have verified the net ferromagnetic moment in the ordered phase of $\rm{Nd_3Sb_3Mg_2O_{14}}$, confirming non-zero scalar chirality and non-zero Berry curvature, leading us to expect topologically protected edge states. Even so, the data clearly indicate a ferromagnetic magnetization and therefore a net scalar chirality from the AIAO structure, leading to the expectation of topological features \cite{Hirschenberger2015,Owerre2017,Laurell_2018}. We have also provided unambiguous evidence of a local magnetic moment reduced to less than 2/3 the expected value in $\rm{Nd_3Sb_3Mg_2O_{14}}$ by measuring nuclear hyperfine excitations, which precludes static disorder as an explanation.

We have also quantified the excitation gap ($38\pm1.4 \> \mu$eV) using specific heat measurements, showing that the ordered magnetic moment reduction cannot be from dynamic spin disorder, leaving the possibility that the reduced moment is due to order in a rotated basis (possibly on the octupolar level). Comparing $\rm{Nd_3Sb_3Mg_2O_{14}}$ to other Nd$^{3+}$ pyrochlores, we have 
argued that the $\rm{Nd_3Sb_3Mg_2O_{14}}$ magnetic Hamiltonian is close to a moment fragmented crystallized monopole state.

\subsection*{Acknowledgments}
This work was supported through the Institute for Quantum Matter at Johns Hopkins University, by the U.S. Department of Energy, Division of Basic Energy Sciences, Grant DE-SC0019331 and by the Gordon and Betty Moore foundation under the EPIQS program GBMF4532.
Use of the NCNR facility was supported in part by the National Science Foundation under Agreement No. DMR-1508249. Thanks also to Oleg Tchernyshov for many helpful discussions.


\begin{thebibliography}{33}
\bibitem [{NIS()}]{NIST_disclaimer}%
	\BibitemOpen
	\href@noop {} {}\bibinfo {note} {Certain commercial instruments are
		identified in this paper to foster understanding. Such identification does
		not imply recommendation or endorsement by the National Institute of
		Standards and Technology, nor does it imply that the instruments identified
		are necessarily the best available for the purpose.}\BibitemShut {Stop}%
\end{thebibliography}

%

\pagebreak

\section*{Supplemental Information for Homogenous reduced moment in a gapful scalar chiral kagome antiferromagnet}

\renewcommand{\thefigure}{S\arabic{figure}}
\renewcommand{\thetable}{S\arabic{table}}
\renewcommand{\theequation}{S.\arabic{equation}}
\renewcommand{\thepage}{S\arabic{page}}  

\setcounter{figure}{0}
\setcounter{page}{1}
\setcounter{equation}{0}

\section{Fitting Hyperfine Excitations}

The nuclear spin incoherent scattering cross section for Nd$^{145}$ is smaller than that of Nd$^{143}$ by an order of magnitude, so we are mostly sensitive to spin flip scattering from Nd$^{143}$. Thus, we use a single Gaussian peak to fit the nuclear hyperfine enhanced inelastic neutron scattering data.

The instrumental resolution is inferred from the variance of the incoherent elastic peak $\sigma_{res} = 0.384(4) \>\mu$eV. After correcting for this resolution we find the physical variance of the inelastic peak to be $\sigma_{mom} = 0.29(5) \>\mu$eV at 0.1 K $\sigma_{mom} = 0.32(6) \>\mu$eV at 0.3 K. With the empirical relation between Nd hyperfine energy and ordered moment (eq. 2), this translates to $\pm 0.23(4) \> \rm \mu_B$ at 0.1 K, $\pm 0.26(5) \> \rm \mu_B$ at 0.3 K as noted in the text.

An alternative to the empirical relation for fitting the hyperfine excitations is calculating the excitation energy directly from a nuclear hyperfine Hamiltonian. We tried this as well, using the hyperfine coupling constants given in ref. \cite{Bleaney1963}. The results from this calculation were $2.17(2) \> \rm \mu_B$ at 0.1 K and $2.10(3) \> \rm \mu_B$ at 0.3 K with $\sigma_{mom} = 0.19(2) \> \rm \mu_B$ and $\sigma_{mom} = 0.23(3) \> \rm \mu_B$ respectively. This still shows a 25\% reduction from the theoretical ordered moment, but the ordered moment is greater than predicted by the empirical formula from ref. \cite{Chatterji2008}. We attribute this discrepancy to the fact that the hyperfine coupling is somewhat sample dependent, and the values in ref. \cite{Bleaney1963} were derived from rare earth metals, whereas the empirical formula in ref. \cite{Chatterji2008} was derived for oxide insulators such as we have here. Thus, because our compound is more similar to those in ref. \cite{Chatterji2008}, we take the values of the empirical formula as more accurate. However, the moment variance from the hyperfine model includes broadening from the two different isotopes, so we take the moment variances from this model to be more accurate.

\section{Spin wave excitations}

\begin{figure}
	\centering\includegraphics[scale=0.55]{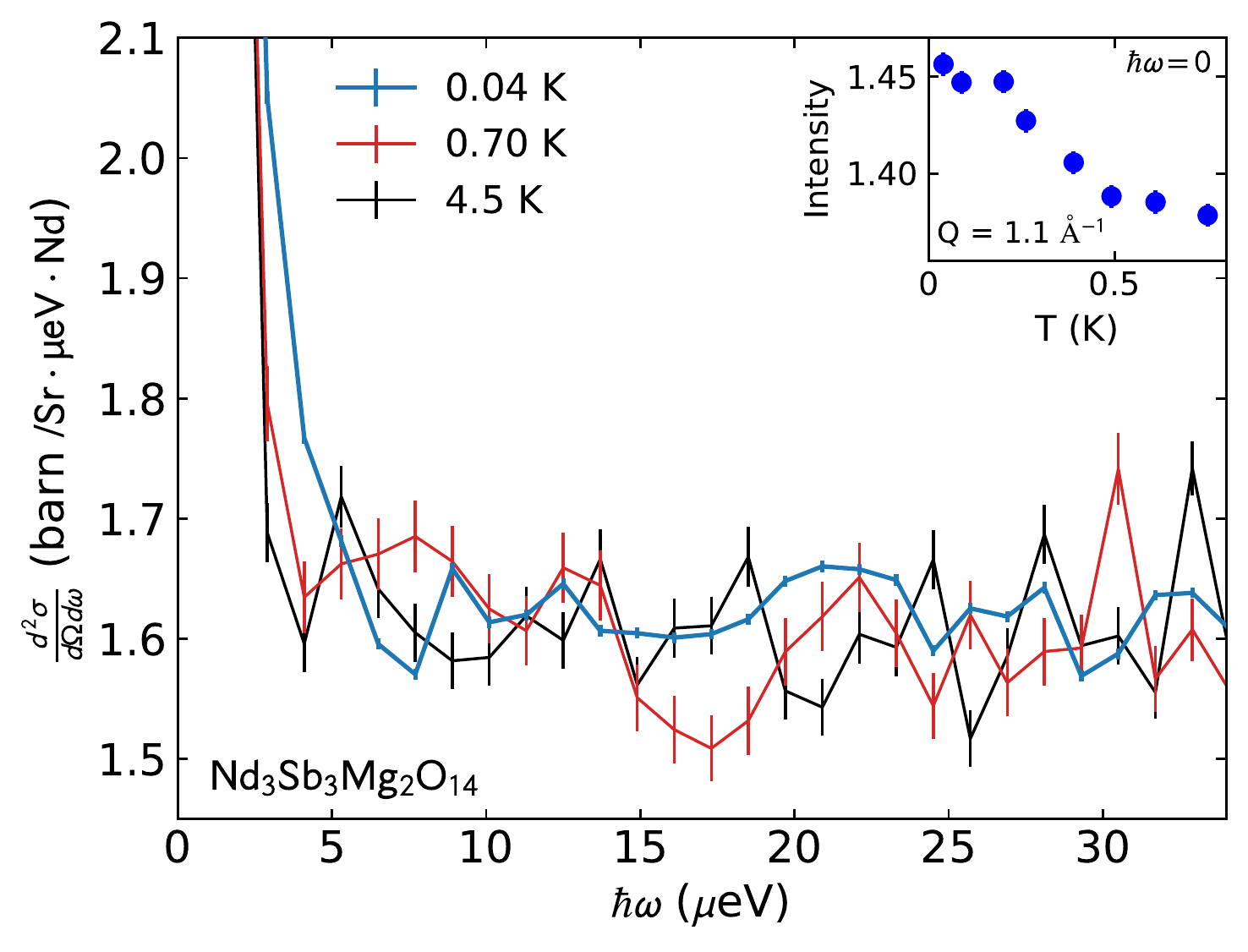}
	
	\caption{Inelastic neutron spectrum of $\rm{Nd_3Sb_3Mg_2O_{14}}$ from 0 $\rm \mu eV$ to 36 $\rm \mu eV$ between $|Q|=1.1$ \AA$^{-1}$ and 1.3 \AA$^{-1}$. No spin wave scattering is apparent, which is consistent with an excitation gap exceeding the measurement bandwidth. The elastic channel (inset) at 1.1 \AA$^{-1}$ clearly shows the onset of magnetic order from the (101) magnetic Bragg peak. Error bars represent one standard deviation.}
	\label{flo:fullSpectrum}
\end{figure}

We initially hoped to measure the spin wave density of states in $\rm{Nd_3Sb_3Mg_2O_{14}}$, but no signal was visible. The neutron backscattering spectrum in $\pm 36 \> {\rm \mu eV}$ mode data are in Fig. \ref{flo:fullSpectrum}. We also collected fixed window (elastic only) data cooling in 100 mK steps to ensure thermal equilibrium conditions during data acquisition. These data are shown in the inset of Fig. \ref{flo:fullSpectrum}. Figure \ref{flo:fullSpectrum} shows the full neutron spectrum for wave vector transfer  between 1.1 \AA$^{-1}$ and 1.3 \AA$^{-1}$, which is near the (101) Bragg peak ($Q(101)=1.041$\AA$^-1$).
Although the sample was clearly in the magnetic ordered phase (Fig. \ref{flo:fullSpectrum} inset), no statistically significant changes were observed  between spectra acquired above and below TN. This is consistent with the $38 \> {\rm \mu eV}$ gap gap in the excitation spectrum inferred from specific heat data. 

\section{Specific Heat Fits}

To quantify the spin wave gap size, we fitted the low temperature heat capacity to Eq. 3 of the main text. This equation was integrated numerically from $q=0$ to $\sqrt{(35 k_B T)^2 + \Delta^2}/c$ (at which point the expression inside the integral is nearly zero) with 5000 steps, and we found that increasing the number of steps or the upper integration bound did not significantly change the results. Therefore we are confident that the numerical routine accurately captures the integral's behavior.

Because the Schottky anomaly calculation in ref. \cite{MyPaper} haa a less accurate estimate of hyperfine splitting, we also recalculated and fit the nuclear Schottky anomaly. The final fitted values were a gap of $\Delta = 38 \pm 1.4 \> \mu{\rm eV}$, a fitted spin wave velocity of $c = 46.31 \pm 0.08 \> {\rm m/s}$, and a fitted ordered moment of $2.06 \pm 0.03 \> {\rm \mu_B}$ (quoted uncertainties are statistical only). 
The fitted spin wave velocity is higher than our previous estimate of 33 m/s  \cite{MyPaper}. This is because of a calculation error in ref. \cite{MyPaper} so the new estimate is the correct one. The ordered moment inferred  from the nuclear Schottky anomaly is also different, but only because the moment reported in \cite{MyPaper} was $\sqrt{\langle J \rangle(\langle J \rangle+1)}$, whereas we quote $\langle J \rangle$.

We also fit the nuclear Schottky anomaly of $\rm Nd_2Zr_2O_7$ (shown in Fig. 4 in the main text), and we found that a Schottky anomaly computed with the measured ordered moment $0.8 \> \rm \mu_B$ dramatically underestimates the observed low temperature upturn in heat capacity. To get the nuclear hyperfine specific heat to match the data, we had to assume an ordered moment close to $2.7 \> \rm \mu_B$. This splitting indicates a static dipolar Nd order in $\rm Nd_2Zr_2O_7$ far larger than what is reported in refs. \cite{Lhotel_2015,Petit2016}. 
Neutron diffraction probes on a shorter timescale than low temperature specific heat, so spin fluctuations cannot account for the discrepancy. This leaves three possibilities: (a) there is static disorder in $\rm Nd_2Zr_2O_7$ which suppresses the ordered moment, (b) there is a strong sample dependence to the magnetic order such that in ref. \cite{Blote1969} the sample is mostly static order and in refs. \cite{Hatnean_2015,Petit2016} the sample is mostly dynamic, and (c) the sample may not have been as cold as indicated by thermometry in refs. \cite{Hatnean_2015,Petit2016}. The hypothesis of moment fragmentation supports possibility (a), where a crystallized monopole state would have a local magnetic ordered moment of twice the average ordered magnetic moment.  Presumably, defects and disorder could also cause the reduced moment to occur, but this remains to be explored experimentally.

\subsection{High temperature expansion of heat capacity}

Here we derive eq. 4 in the main text using a high temperature expansion of specific heat. Given a dispersion of the form $\epsilon(q) = \sqrt{c^{2}q^{2} + \Delta^{2}}$ in three dimensions, $\Delta$ being the energy gap, we have the relation $\Delta \propto \sqrt{x_i}$ where $x_i$ is x the intercept on the $\frac{C}{T}$ vs $T^{2}$ graph.

The energy associated with bosonic spin wave excitations is given by
\begin{equation}
\label{eq.U(T)_q}
u(T) = \frac{1}{2\pi^{2}}\int_{0}^{\infty}\frac{\epsilon(q)q^{2}}{e^{\beta\epsilon(q)}-1}dq
\end{equation}
\noindent
where $\beta = 1/(k_{B}T)$. Introducing the density of states $g(\epsilon) = \frac{V}{2\pi^{2}}q^{2}\frac{\partial q}{\partial \epsilon} = \frac{V}{2c^{3}\pi^{2}}\epsilon\sqrt{\epsilon^{2}-\Delta^{2}}$, then $u(T)$ can be rewritten as 

\begin{equation}
\label{eq.U(T)_e}
u(T) = \frac{1}{2\pi^{2}}\int_{\Delta}^{\infty}\frac{\epsilon g(\epsilon)}{e^{\beta\epsilon}-1}d\epsilon = \frac{1}{2c^{3}\pi^{2}}\int_{\Delta}^{\infty}\frac{\epsilon^{2}\sqrt{\epsilon^{2}-\Delta^{2}}}{e^{\beta\epsilon}-1}d\epsilon.
\end{equation}
In the integral we use the substitution $z = \beta\epsilon$, to obtain an integral in terms of $z$,
\begin{equation}
\label{eq.U(T)_z}
u(T) = \frac{1}{2c^{3}\pi^{2}}\frac{1}{\beta^{4}}\int_{|\beta\Delta|}^{\infty}\frac{z^{2}\sqrt{z^{2}-(\beta\Delta)^{2}}}{e^{z}-1}dz.
\end{equation}
Let us concentrate on the functional form of the integral $f(\beta\Delta) =\int_{|\beta\Delta|}^{\infty}\frac{z^{2}\sqrt{z^{2}-(\beta\Delta)^{2}}}{e^{z}-1}dz$. From the structure of the integral we infer that $f(\beta\Delta)$ is even, and that it is non singular at $\beta\Delta = 0$. In fact $f(0) = \frac{\pi^{4}}{15}$. Hence we can always expand $f$ as a power series in $\Delta\beta$ around $\Delta\beta = 0$ and drop higher order terms in a high temperature (small $\beta$) approximation. This gives us an expression for the energy as:
\begin{equation}
\label{eq.U_series}
u(T) =  \frac{1}{2c^{3}\pi^{2}}\sum_{n = 0}^{n = \infty}\alpha_{n}\Delta^{n}T^{4-n}.
\end{equation}
From this we calculate the specific heat as,
\begin{eqnarray}
\frac{c_{v}}{T} &=& \frac{1}{T}\frac{\partial u}{\partial T} = \frac{1}{2c^{3}\pi^{2}}\sum_{n = 0}^{n = \infty}(4-n)\alpha_{n}\Delta^{n}T^{2-n}. \\ \nonumber
\frac{c_{v}}{T} &=& 4\alpha_{0} T^{2} + 3\alpha_{1} \Delta T + 2\alpha_{2} \Delta^{2} + \mathcal{O}\left(\frac{1}{T}\right).
\end{eqnarray}
In the series expansion at high T we can drop terms of order $\mathcal{O}\left(\frac{1}{T}\right)$. The quadratic equation can now be solved to obtain an expression for the intercept on the $T^{2}$ axis. Setting $4\alpha_{0} T^{2} + 3\alpha_{1} \Delta T + 2\alpha_{2} \Delta^{2} = 0$ we obtain $T = \frac{\Delta}{8\alpha_{0}}(-3\alpha_{1} \pm \sqrt{9\alpha_{1}^{2} - 32\alpha_{0}\alpha_{2}})$ as the solution. Notably we have $T \propto \Delta$ and hence if our x-axis is $T^{2}$ we have the relation $\Delta \propto \sqrt{x_{i}}$, where $x_i$ is the $x$ intercept.

Let us now extract the proportionality constant from the original integral. To do this we make two approximations. Firstly, we binomially expand the numerator of the integral Eq.(\ref{eq.U(T)_z}) as $z^{3}\left(1-\frac{(\Delta\beta)^{2}}{2z^{2}}\right)$, retaining the lowest order correction term only. Note the limits of the integral impose $\frac{\Delta\beta}{z} < 1$ and the next term in the series is quartic in $\beta$ which is a small number. Secondly we extend the integration limits from $\int_{|\beta\Delta|}^{\infty} \to \int_{0}^{\infty}$. This assumption neglects terms of order $|\beta\Delta|^{3}$ and higher. It is justified in hindsight as the coefficient obtained with it matches numerical simulations quite well. We are thus left with,
\begin{eqnarray}
\nonumber
u(T) &=& \frac{1}{2c^{3}\pi^{2}} \frac{1}{\beta^{4}} \left( \int_{0}^{\infty} \frac{z^3}{e^z - 1}dz - \frac{(\beta\Delta)^{2}}{2}\int_{0}^{\infty}\frac{z}{e^z - 1}dz \right). \\ \nonumber
u(T) &=& \frac{1}{2c^{3}\pi^{2}} \frac{1}{\beta^{4}} \left( \frac{\pi^{4}}{15} - \frac{\pi^2}{12}(\beta\Delta)^{2} \right).
\end{eqnarray}
Using this expression we obtain the specific heat as $c_{v} = \frac{1}{2c^{3}\pi^{2}}\left(\frac{4k_{B}^{4}\pi^{4}}{15}T^{3} - \frac{2k_{B}^{2}\pi^2}{12}\Delta^{2} T \right)$. This we can solve for the intercept $x_i$ as,
\begin{equation}
\Delta = \left(\sqrt{\frac{8}{5}}\pi k_{B}\right)\sqrt{x_{i}}.
\end{equation}

\section{Magnetization}

\begin{figure}
	\centering\includegraphics[scale=0.48]{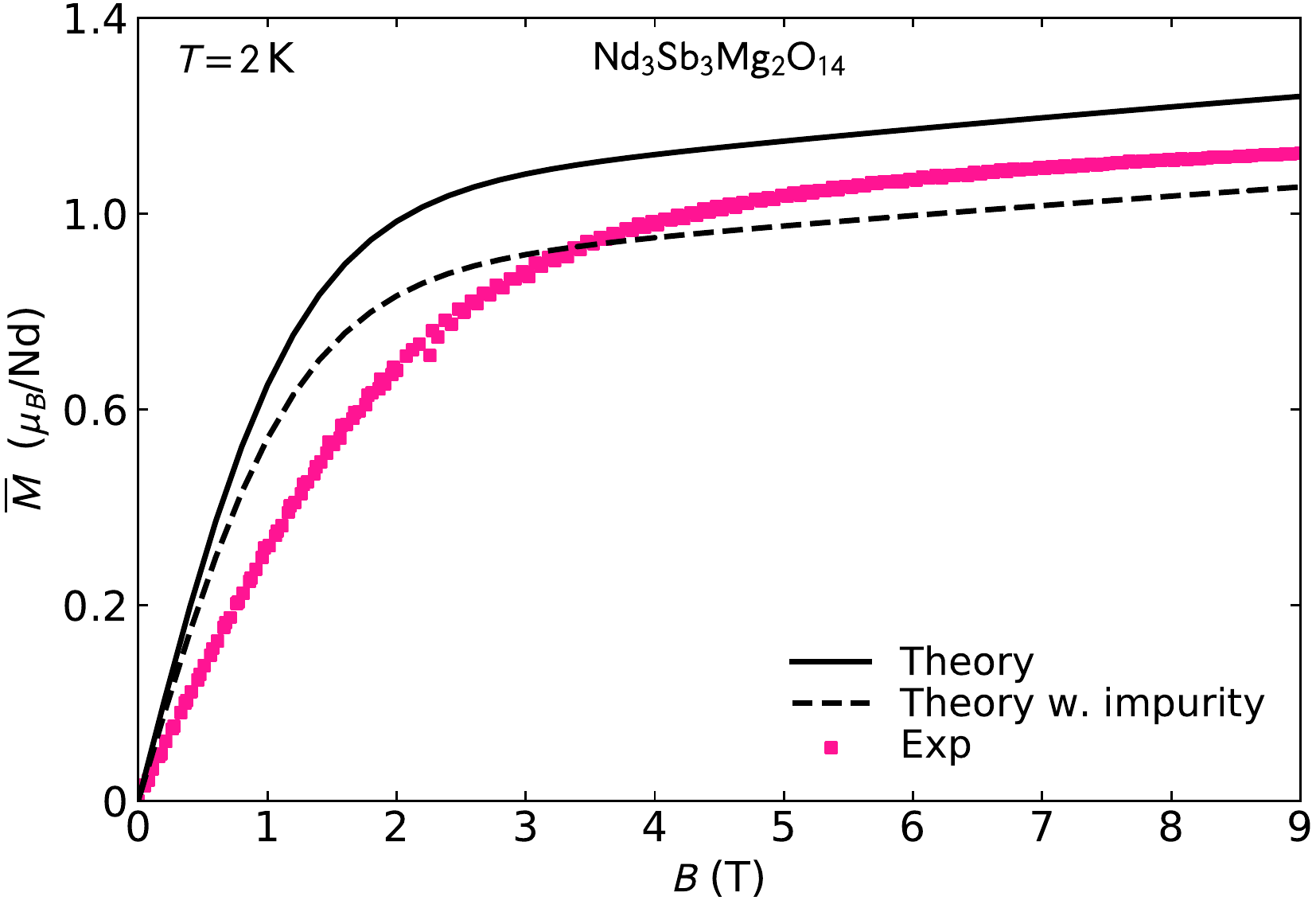}
	
	\caption{Powder-averaged magnetization of $\rm{Nd_3Sb_3Mg_2O_{14}}$ from 0 T to 9 T at 2 K. The measured value is smaller than the calculated magnetization from the CEF Hamiltonian (solid line), but larger than the calculated value assuming 13\% nonmagnetic impurities (dashed line). Given the presence of magnetic impurities not accounted for by this calculation  \cite{Scheie2018_CEF}, it is reasonable to expect that the measured saturation magnetization lies somewhere between these two values.}
	\label{flo:Satmagnetization}
\end{figure}

Because the $\rm{Nd_3Sb_3Mg_2O_{14}}$ zero-field ordered moment is reduced, it is worthwhile to ask whether the moment extracted from saturation magnetization is similarly reduced. Simulating the magnetization using PyCrystalField \cite{PyCrystalField} and the Hamiltonian derived in ref. \cite{Scheie2018_CEF}, we find that the estimated saturation magnetization is larger than measured in experiment (data from ref. \cite{MyPaper}) as shown in Fig. \ref{flo:Satmagnetization}.

The calculated powder-averaged saturated moment is larger than measured, but this is reconciled when we assume 13\% impurities (inferred from susceptibility in ref. \cite{Scheie2018_CEF}). We do not know the the magnetization curve of the impurities, so we plotted the Nd CEF result assuming nonmagnetic impurities with a dashed line. The experimental data lies between these two predictions, which is reasonable for an impurity contribution.
The differences in initial slope in magnetization are probably due to magnetic exchange \cite{MyPaper}, which is not accounted for in this calculation.

\section{Octupolar pseudospin components}

The ground state doublet reported in ref. \cite{Scheie2018_CEF} is very Ising-like, but the symmetry of the environment is so low that there are many pathways for transverse terms to appear---particularly if octupolar exchange is present (as is theorized for other Nd pyrochlores). These pathways may stabilize magnetic order in a basis which includes octupolar order.

The most general equation for exchange between two spins is
\begin{equation}
H = \sum_{\Lambda,\Lambda'}\sum_{\mu,\mu'} I^{\mu\mu'}_{\Lambda\Lambda'} O_{\Lambda}^{\mu} O_{\Lambda'}^{\mu'}
\end{equation}
where $\Lambda$ is the order of the multipole (1 = dipole, 2 = quadrupole, 3 = octupole, etc.), $\mu = -\Lambda, ... , \Lambda$, $O_{\Lambda}^{\mu}$ are Stevens Operators, and $I$ are the exchange constants \cite{Santini_2009}. One can simplify the expression by defining pseudospin vectors based off the ground state eigenkets of an ion with
\begin{equation}
\tau_{\Lambda} = \sum_{\mu} \langle \sigma' | O_{\Lambda}^{\mu} | \sigma \rangle
\end{equation}
where $| \sigma \rangle$ and $| \sigma' \rangle$ are the ground state doublet, yielding a set of $\tau_{\Lambda}$ which are $2\times 2$ Pauli spin matrices \cite{Hermele_2014,Li_2016,Benton_2016}. The pseudospin operators up to rank 3 for $\rm{Nd_3Sb_3Mg_2O_{14}}$, based of the CEF ground state determined in ref. \cite{Scheie2018_CEF}, are
\begin{equation}
\tau_1=
\left( {\begin{array}{cc}
	-3.469 & 0.479-0.231i \\
	0.479+0.231i & 3.469 \\
	\end{array} } \right)
\end{equation}
\begin{equation}
\tau_2=
\left( {\begin{array}{cc}
	23.47 & 0.0 \\
	0.0 & 23.47 \\
	\end{array} } \right)
\end{equation}
\begin{equation}
\tau_3=
\left( {\begin{array}{cc}
	-89.94 & -52.14-6.99i \\
	-52.14+6.99i & 89.94 \\
	\end{array} } \right).
\label{eq:octupolarpseudospin}
\end{equation}
The diagonal elements represent Ising components $\tau_z$, and the off-diagonal elements represent transverse components $\tau_{\pm}$. The dipolar pseudospins are clearly Ising-like, with the transverse components an order of magnitude smaller than the $z$ components. The same holds for the quadrupolar case. The octupolar pseudospin, however, has significant transverse and longitudinal components, which could be a mechanism for fluctuating spins if octupolar exchange coupling is strong.

\end{document}